\documentclass[10pt]{article}
\usepackage{multirow}
\usepackage[dvips]{graphicx}

\usepackage[psamsfonts]{amssymb}
\usepackage{amsxtra}
\usepackage{threeparttable}
\usepackage{amsmath}
\usepackage{amssymb}

\usepackage{graphicx}

\usepackage{cite}

\usepackage{color}

\usepackage{setspace}
\usepackage{float}
\usepackage{subfigure}

\usepackage[labelfont=bf,labelsep=period,justification=raggedright]{caption}

\makeatletter
\renewcommand{\@biblabel}[1]{\quad#1.}
\makeatother

\date{}

\begin{document}

\begin{flushleft}
{\Large
\textbf{The Spatial Real and Virtual Sound Stimuli Optimization for the Auditory BCI}
}

Nozomu Nishikawa$^{1}$,
Yoshihiro Matsumoto$^{1}$, 
Shoji Makino$^{1}$ 
and
Tomasz M. Rutkowski$^{1,2,\ast}$
\\
\bf{1} Multimedia Laboratory, TARA Center, University of Tsukuba, Tsukuba, Japan\\
\bf{2} RIKEN Brain Science Institute, Wako-shi, Japan\\
$\ast$ E-mail: tomek@tara.tsukuba.ac.jp
\end{flushleft}

\section*{Abstract}

The paper presents results from a project aiming to create horizontally distributed surround sound sources and virtual sound images as auditory BCI (aBCI) stimuli. The purpose is to create evoked brain wave response patterns depending on attended or ignored sound directions. 
We propose to use a modified version of the vector based amplitude panning (VBAP) approach to achieve the goal. The so created spatial sound stimulus system for the novel oddball aBCI paradigm allows us to create a multi--command experimental environment with very encouraging results reported in this paper. We also present results showing that a modulation of the sound image depth changes also the subject responses. Finally, we also compare the proposed virtual sound approach with the traditional one based on real sound sources generated from the real loudspeaker directions. The so obtained results confirm the hypothesis of the possibility to modulate independently the brain responses to spatial types and depths of sound sources which allows for the development of the novel multi-command aBCI.

\section{Introduction}

In recent years, many researchers who study the brain waves plug to the technological development of using electroencephalography (EEG) signals for engineering applications. The research field has made a remarkable development. 
It is a fact that thanks to such application the disabled people could move successfully a cursor or a wheelchair using their brain wave patterns only. 
Thanks to the BCI technology it is expected that disabled people will receive the various services only by thinking or interacting with their environments~\cite{bciBOOKwolpaw}.

\begin{figure}[H]
\begin{center}
\includegraphics[width=0.7\linewidth]{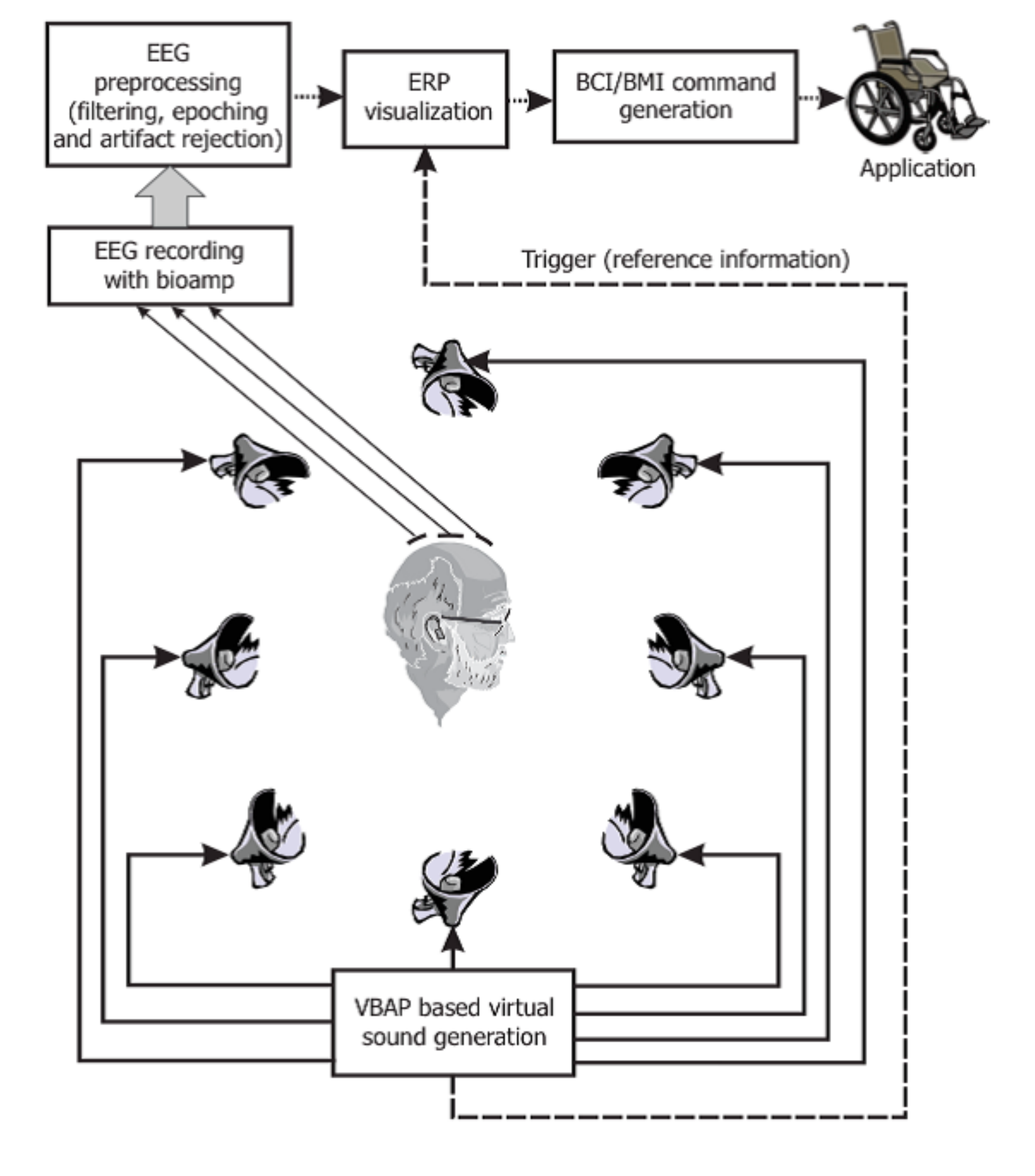}
\end{center}
\caption{Schematic drawing of the proposed auditory BCI system. The subject in the center is surrounded by horizontally distributed loudspeakers. EEG signals are processed by the BCI processing stages in synchrony with presented spatial auditory stimuli.}
\label{fig:AuditoryBCI}
\end{figure}

The spatial auditory BCI (saBCI) is the interface that makes use of the brain waves captured in response to spatial sound stimuli.
In our research, we aim at the practical realization and enhancement of the saBCI~\cite{nozomuB4thesis}.
Figure~\ref{fig:AuditoryBCI} depicts our saBCI system prototype that performs the following: 
\begin{itemize}
	\item It creates the spatial sound stimuli in the octagonal surround sound loudspeaker system using the modified VBAP approach as well as the conventional real sound sources;
	\item It captures and analyzes the EEG event related responses (ERP) to spatial sound stimuli;
	\item Finally it classifies the responses in order to generate the BCI application command (exemplary wheelchair in the diagram). This paper does not include yet classification results.
\end{itemize}

\begin{figure}[H]
\begin{center}
\includegraphics[width=0.8\linewidth]{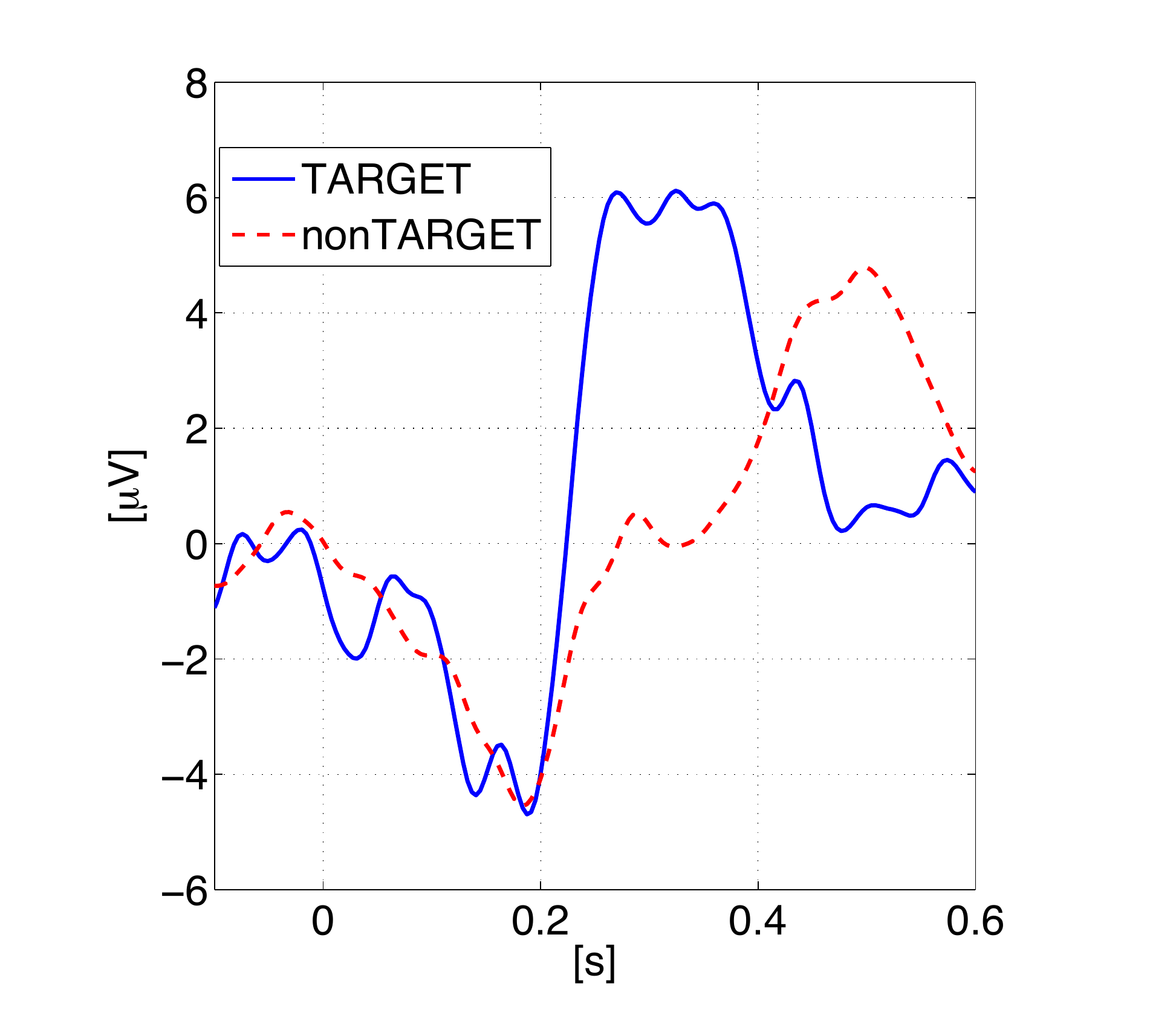}
\end{center}
\caption{An example of averaged brain evoked responses to \emph{target} (solid blue line) and \emph{non-target} (dotted red line). The difference in ERP waves is mostly pronounced around $300$ms where usually $P300$\emph{/aha-response} occurs. The zero second time point indicates stimulus onset.}
\label{fig:P300}
\end{figure}

In the presented project we refer to the fact that a human brain given a sound stimulus generates the electric ERP pattern~\cite{book:eeg}. Attentional modulation adds a characteristic deflection (so called ``aha response'') which could be utilized in the saBCI paradigm development. This ERP feature depending on whether the sound is attended (so called \emph{target}) or not (\emph{non-target}) allows for further classification of subject's choices~\cite{tomekHAID2011,tomekAPSIPA2011}.
``The aha-response'' to the attended \emph{target} sound in comparison with the other \emph{non-target} has the characteristic positive deflection after $300$ms from the onset and thus it is called ``the $P300$ response.'' An example of an averaged $P300$ response (solid blue line - \emph{target}), from our experiments, is depicted in Figure~\ref{fig:P300} together with the ignored sound direction (dashed red line - \emph{non-target}).
The properly classified, by a machine learning application, $P300$ response lets the BCI application user to select a command. In this paper we only identify the responses separability and the future research will result with final classification.

There are multiple optimization steps in a course of the saBCI system development. The first step is an optimal choice of the number of loudspeakers in order to create a multi--command system. In the state-of-the-are saBCI applications~\cite{tomekAPSIPA2011,bciSPATIALaudio2010} the number of commands depends on the number of real loudspeakers, since they generate the sound sources. We aim to increase the number by employing a concept of a virtual sound sources or so called virtual sound images in our saBCI prototype. This will allow to decrease the number of the necessary loudspeakers or to increase a number of commands in the current setups.
The virtual sound image is generated using multiple loudspeakers, not with a single real one. The VBAP method to generate virtual sound images is described in the following sections together with the proposed multi-command saBCI application. Additionally we propose to utilize sound depth panning to further possibly increase a number of commands in a limited loudspeaker number environment. 

We conduct a series of psychophysical and offline saBCI EEG experiments in order to address the following questions: 
\begin{enumerate}
	\item Does the spatial sound stimuli delivery system influences the results? In detail, is there any significant difference between real and virtual sound images?
	\item Does a type of sound stimuli timbre matter?
	\item Does the sound image depth (modulated volume in our case) influences the responses?
\end{enumerate}

The paper from now on is organized as follows. First we introduce spatial sound creation and delivery system. Next psychophysical and offline asBCI experiments are described. Finally results discussion concludes the paper.

\section{Methods}

This section describes the stimuli generation methods, EEG capture experimental protocol in offline saBCI mode, and the ERP processing steps in order to elucidate and prove separability of the brain $P300$ responses for the future online multi-command application with a classifier.

\subsection{VBAP Method for Generating of Spatial Virtual Sound Sources in Application to saBCI}

In the proposed saBCI paradigm the user receives the real and virtual sound source stimuli that are created using a \textsf{MAX} environment~\cite{maxMSP}. The surround sound environment consist of the eight loudspeakers distributed octagonally at a single meter distance from the subject's head as depicted in Figure~\ref{fig:AuditoryBCI}. 

\begin{figure}[H]
\begin{center}
\includegraphics[width=0.65\linewidth]{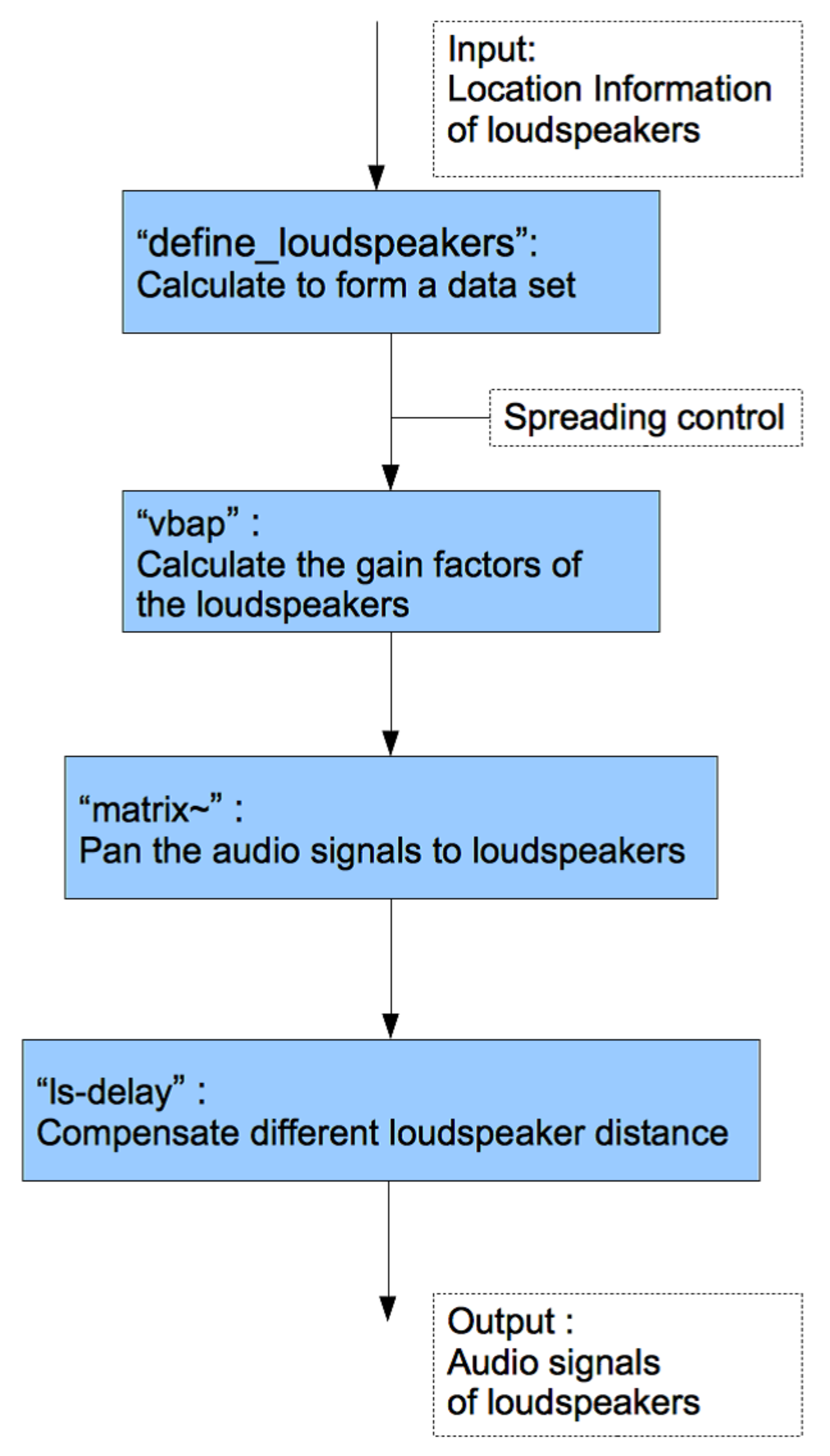}
\end{center}
\caption{A schematic figure of the VBAP method implementation.}\label{fig:vbap}
\end{figure}

We implement a custom VBAP~\cite{pulkkiVBAP1997} patch in \textsf{MAX} environment that controls the eight spatially distributed loudspeakers in real time. The developed VBAP patch can be used to position a virtual sound source with any arbitrary loudspeaker configuration. The amplitude panning concept implemented in the VBAP method is formulated with vectors pointing the desired sound source position in the sound field. In our particular configuration of a horizontal plane defined by the octagonally distributed loudspeakers the virtual sound images could be created only the pairs of neighboring loudspeakers. The simplified pair-wise VBAP application reduces the sound amplitude panning to only two loudspeakers used at a time.
The proposed VBAP reformulation leads to simpler equations for amplitude panning, and the use of the vectors makes the panning computationally efficient.
The VBAP implementation consists of three objects (functions) in \textsf{MAX} environment: {\it define\_loudspeakers, vbap}, and {\it matrix$^\sim$}~, as depicted in a flow digram in Figure \ref{fig:vbap}.

\begin{figure}[H]
\begin{center}
\includegraphics[width=0.8\linewidth]{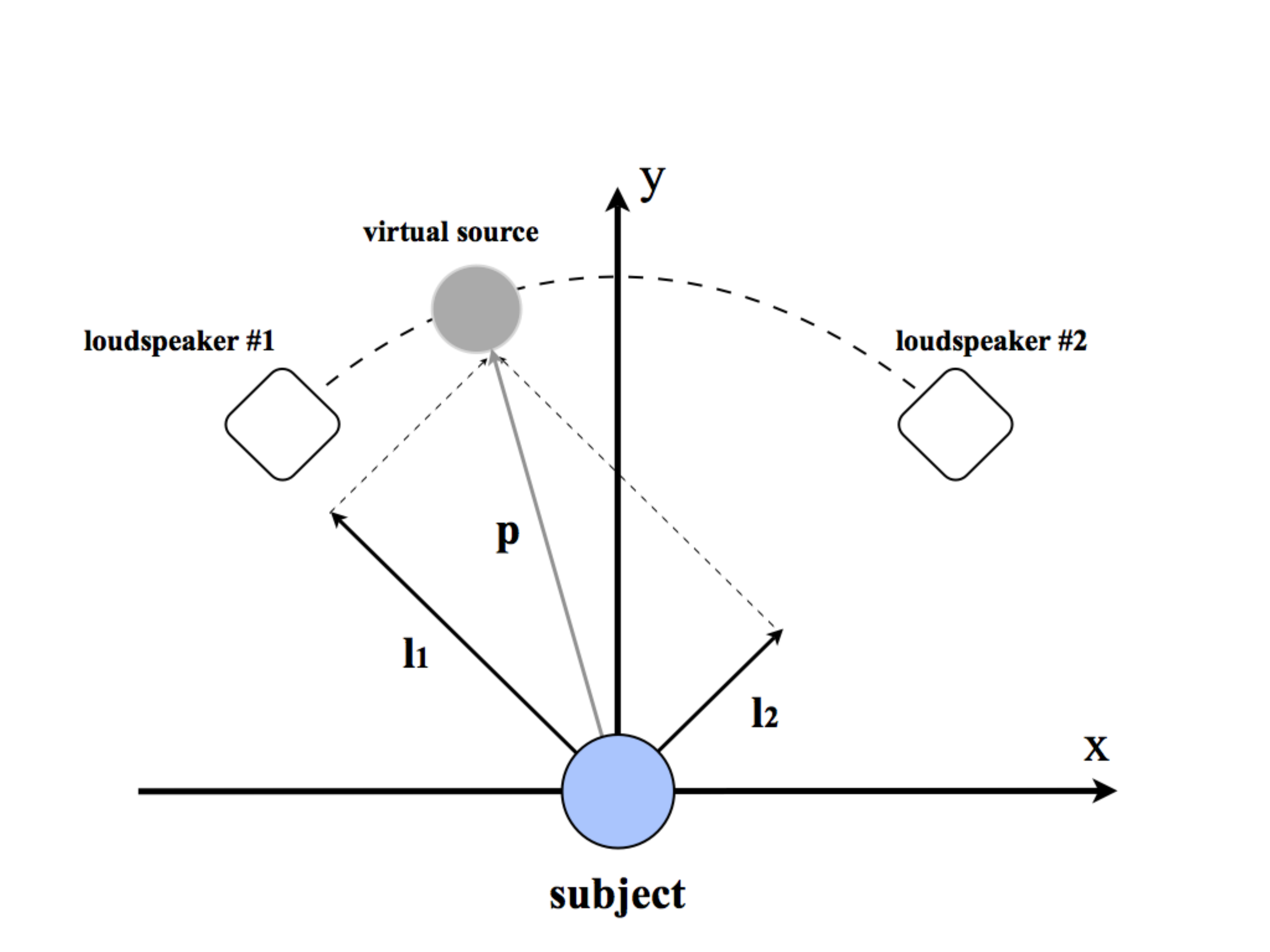}
\end{center}
\caption{A concept of the simplified horizontal panning of sound source with two loudspeakers based on the original VBAP~\cite{pulkkiVBAP1997} implementation.}
\label{fig:2ch_vbap}
\end{figure}

In our proposal of the simplified two-dimensional VBAP method the two-channel loudspeaker configuration is reformulated as a two-dimensional vector base. The base is defined by unit length vectors 
\begin{equation}
	\mathbf{l}_{1} = [l_{1x}~l_{1y}]^{T} \quad \mbox{and} \quad \mathbf{l}_{2} = [l_{2x}~l_{2y}]^{T},\label{eq:speakVECTOR}
\end{equation}
which are pointing toward $the~loudspeaker~\#1$ and $\#2$, respectively, as depicted in Figure~\ref{fig:2ch_vbap}.
The superscript $^T$ denotes the matrix transposition. The $l_{ix}$ and $l_{iy}$ are the $\mathbf{l}_i$ vector coordinates on $x$ and $y$ axes, respectively.  The unit-length vector
\begin{equation}
	\mathbf{p} = [p_x~p_y]^{T}, \label{eq:virtVECTOR}
\end{equation}
pointing toward the desired virtual sound source can be treated as a linear combination of the two neighboring loudspeaker vectors as follows
\begin{eqnarray}
	\mathbf{p} = g_{1}\mathbf{l}_1 + g_{2}\mathbf{l}_2, 
\label{eq:p1}
\end{eqnarray}
where $g_{1}$ and $g_{2}$ are the gain factors, which are the nonnegative scalar variables to be optimized by VBAP in order to position the virtual sound image. The equation (\ref{eq:p1}) could be reformulated in a matrix form as
\begin{eqnarray}
	\mathbf{p} = \left(\mathbf{gL}_{12}\right)^{T},\label{eq:VBAP}
\end{eqnarray} 
where $\mathbf{g} = [g_{1}~g_{2}]$ and $\mbox{\boldmath $L_{12} = [l_{1}~~~l_{2}]$}^{T}$.
The equation (\ref{eq:VBAP}) can be solved if \mbox{\boldmath $L_{12}^{-1}$} exists, so the gain vectors could be obtained as follows
\begin{eqnarray}
	\mathbf{g} = \mathbf{p}^{T}\mathbf{L}_{12}^{-1} = [p_{1}~p_{2}]
\left[
\begin{array}{cc}
l_{1x} & l_{1y} \\
l_{2x} & l_{2y} \\
\end{array}
\right]^{-1}.
\label{eq:g1}
\end{eqnarray}
The inverse matrix $\mathbf{L}_{12}^{-1}$ satisfies $\mathbf{L}_{12}\mathbf{L}_{12}^{-1} = \mathbf{I}$, where $\mathbf{I}$ is the identity matrix. The
gain factors $g_{1}$ and $g_{2}$ calculated using equation~(\ref{eq:g1}).

\subsection{Increasing the Number of the saBCI Commands with the Spatial Depth Adjustment of the Sound Sources}

Until now we only discussed the spatial directional sound sources manipulation in order to create the multi-command saBCI paradigm. Human brain auditory system can also distinguish easily a change of sound source distance by comparing their loudnesses. We propose to test additionally how the changing sound distance or depth could help us to increase a number of possible saBCI commands. In the previously introduced VBAP method in equation~(\ref{eq:VBAP}) the loudspeaker sound amplitudes are controlled with gain factors $g_{1}$ and $g_{2}$, respectively. In traditional VBAP implementation the sum of squares of the gain factors shall always obey 
\begin{equation}
	g_{1}^2+g_{2}^2=1.
\end{equation}	
However, the total sound loudness could be set to a constant value $C$, that would reflect changing sound source depth location as follows
\begin{equation}
	g_1^2 + g_2^2 = c \quad 0\leqslant c \leqslant 1.\label{eq:C} 
\end{equation}
So a depth setting for $c=1$ could be referred as a standard/reference setting. The smaller value of $c$, as in equation~(\ref{eq:C}), will create a farther distance sound illusion. In our experiments discussed in the following sections we use settings of $c=\{0.2, 1\}$ to create sound sources with loudnesses of approximately $\sim30$dB or $\sim70$dB, respectively, in our laboratory's soundproof studio.

\section{The Psychophysical Experiment with Eight Loudspeakers to Evaluate Spatial Sound Stimuli}

%\subsection{Purpose of the psychophysical experiment}

Sound localization performed by human subjects is limited by some features related to head and ear shapes. The most common localization problem is related to front/back sound source confusion, and consequently it is difficult for people to discriminate the exactly frontal or rear positions. 
In order to create and test the auditory stimuli for their difficulty levels we perform first the psychophysical experiment to measure the behavioral responses before the final EEG experiment in offline BCI setting. The spatial target directions resulting in later behavioral responses in form of button presses shall be classified as the more difficult or causing more cognitive load~\cite{book:eeg}. Also a number of mistakenly recognized targets in a random series in each trial identifies possible variable difficulty of the task. Our goal is to design a final spatial oddball paradigm which would require the same cognitive load for all directions in order to have similar $P300$ response easy to classify by machine learning algorithms later.

\begin{figure}[H]
	\begin{center}
	\includegraphics[width=0.8\linewidth]{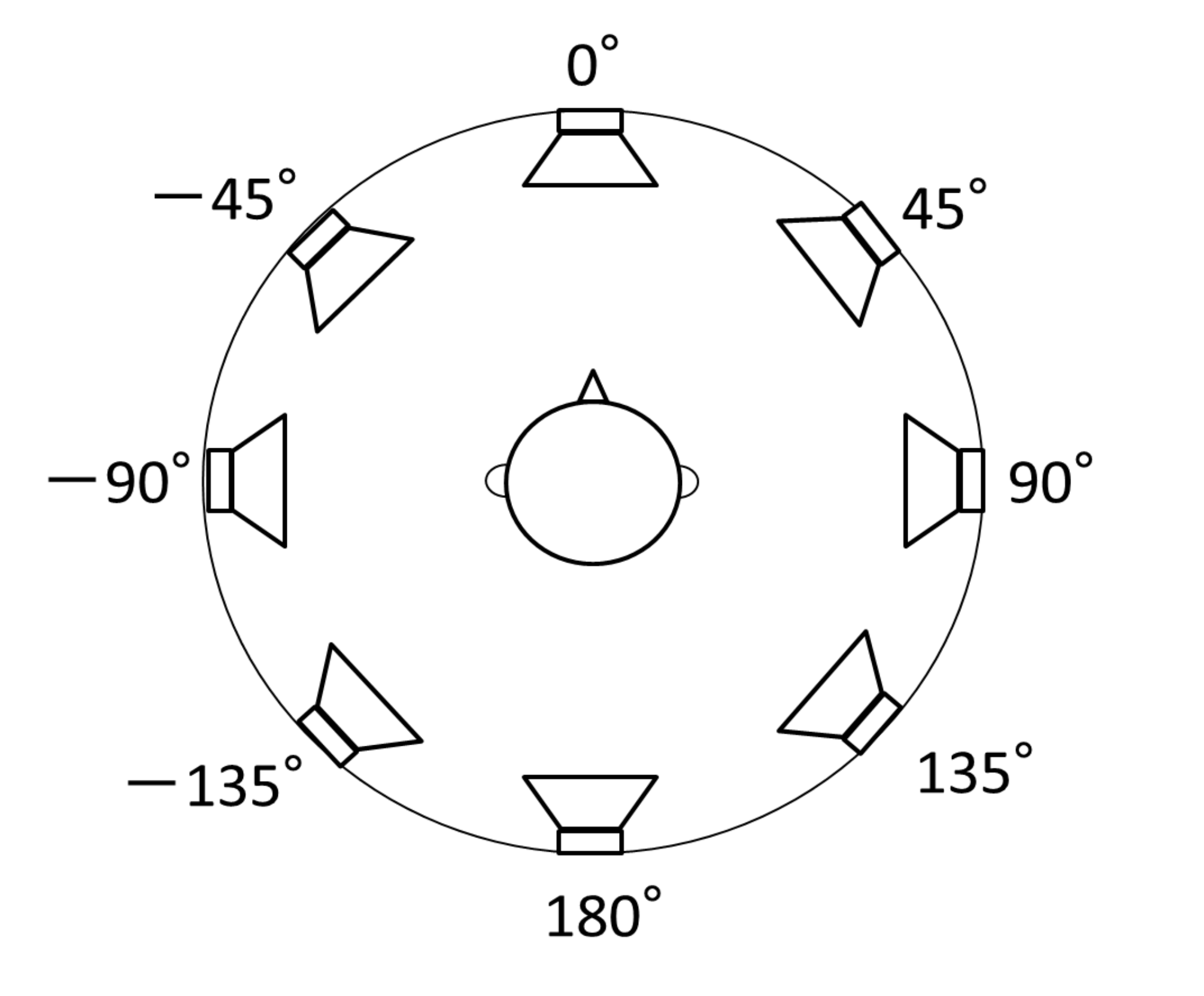}
	\end{center}
	\caption{The locations of the eight loudspeakers in our experiments. The frontal to the subject loudspeaker at a position of $0^\circ$.}
	\label{fig:SpeakerLocation}
\end{figure}

In the series psychophysical experiments described below we compare the behavioral reaction times as modulated by spatial cognitive load difficulty in function of:
\begin{itemize}
	\item Various timbres of the sound stimuli;
	\item Real or virtual sound images creation techniques (real loudspeakers  or VBAP);
	\item Various sound depth settings.
\end{itemize}
In all psychophysical experiments the sound sources are generated with the eight loudspeakers of our system as depicted in Figure~\ref{fig:SpeakerLocation}.

The experiments were conducted with agreement of the local institutional ethical committee guidelines. The subjects were explained the psychophysical and EEG experimental procedures in detail. All of them participated voluntarily. 

In the first experimental session, all the sound sources are generated from the real loudspeaker directions, thus this session is labeled as ``the real.'' The second session utilizes the proposed simplified version of VBAP, as in equation~(\ref{eq:VBAP}), in order to create virtual sound sources at the positions between the neighboring loudspeakers. This second experimental session is labeled as ``the virtual.''
For example in the virtual setting, a sound from a direction of $45^\circ$ is created actually with the two loudspeakers placed at $0^\circ$ and $90^\circ$, instead of the actual one located there in the real experiment. Such setting allows us to compare the real and virtual sound reproduction setting for exactly the same spatial directions.

\subsection{Psychophysical Experiment Protocol}

The procedure of the psychophysical experiment in a single trial is composed of the following steps: 
\begin{enumerate}
	\item An instruction is given to the subject of which spatial sound stimuli direction and which source depth setting (loudness) to attended in form of a sound delivered and followed by a break.
	\item Next the subject listens to the $16$ randomly ordered directions with the same stimulus timbre played from each (each direction has two depth settings).
	\item The subject responses immediately (as fast as possible) by pressing a computer keyboard button after \emph{the target} direction sounds occurs in the sequence, as instructed in the above step 1).
	\item The above procedure is repeated for all the $16$~\emph{target} directions - with single one in each trial.
\end{enumerate}

In order not to overlap the psychophysical responses (usually around $500$ms for our subjects) with the following stimuli the stimuli onset asynchrony (SOA) is set to $1000$ms with stimuli lasting for only the first $500$ms.
The followings are the experimental method to determine the difficulty of each spatial sound stimuli.
After each trial consisting of $16$ spatial sound sources (chance level $= 1/16 = 6.25\%$) we evaluate subject's response correctness and the delay time. Also a comparison among the directions and depth levels is carried.

Table~\ref{tb:BPCondition} summarizes the details of the psychophysical experiment settings. 

We conduct the experiments for the three kinds of stimulus timbre in order to evaluate also the possible subject preferences. The results are discussed in the next section.

\subsection{Psychophysical Experiment Results}

The results of conducted psychophysical experiments are presented in Tables~\ref{tb:REAL_LOU_sti}--\ref{tb:VIR_SML_dir}.
The analysis of sound depths variability, summarized in Tables~\ref{tb:REAL_LOU_sti},~\ref{tb:REAL_SML_sti},~\ref{tb:VIR_LOU_sti}, and~\ref{tb:VIR_SML_sti}, resulted with the lowest and no significant differences of response times, as tested with \emph{Wilcoxon-pariwise-test} for the means. This result  supports the hypothesis of the possibility to utilize this feature for the saBCI paradigm design in real and virtual sound reproduction settings.

On the other hand the results summarized in Tables~\ref{tb:REAL_LOU_dir},~\ref{tb:REAL_SML_dir}, for real sound sources, and in Tables~\ref{tb:VIR_LOU_dir},~\ref{tb:VIR_SML_dir} for virtual ones, resulted also in no significant differences for directions based on the response time analyzes. 

However a comparison of accuracy rates (AR) for real and virtual sound sources in all Tables~\ref{tb:REAL_LOU_sti}--\ref{tb:VIR_SML_dir} showed significant differences with much lower values for virtual. This shows that VBAP application still needs further research on the sound quality related to virtual image spatial spreads. The differences show that virtual sound images are more difficult to localize so they shall not be mixed with the real sounds in a single experimental trials in the offline saBCI experiments as presented in the next section.

\section{The Offline saBCI Experiment}

%\subsection{Purpose of low-density EEG experiments}

Based on previously discussed very encouraging psychophysical experiment results showing no relation between subjects behavioral performance and spatial sound directions we proceed to the offline saBCI paradigm EEG recordings in which the brain responses will be measured to further test our original hypothesis of the possibility to utilize real and proposed virtual spatial sound sources in the oddball experimental paradigm. Now we verify whether the ERP responses show any differences for directions and reproduction system types. Since the previously conducted psychophysical experiment indicated no relation to sound timbres, in the following analysis of EEG responses we average all types.
In order to do so we conduct the following experiments with the same six volunteers and with accordance of the institutional ethical committee guidelines.

The EEG signals are captured with eight dry electrode portable wireless system by \textsf{g.tec} (\textsf{g.MOBIlab+} \& \textsf{g.SAHARA}). The EEG signals are sampled at $256$Hz rate and stored on a laptop computer running custom \textsf{MATLAB Simulink} patch. The user interface and stimulus are created in \textsf{MAX}.

\subsection{The Offline saBCI EEG Experimental Protocol}

The offline saBCI experimental protocol for a single trial is as follows: 
\begin{enumerate}
	\item First an auditory instruction is given to the subject to which spatial sound stimuli to attended.
	\item Next the subject listens to the $16$ randomly ordered stimuli, but this time no any behavioral response is given and only counting of targets is advised to keep a focused attention.
	\item The above procedure is repeated for all \emph{target} directions and volumes separately, resulting in $16$ trials for each timbre.
\end{enumerate}
The detailed experimental setting are summarized in Table~\ref{tb:Final_Condition}.

\subsection{EEG Signal Preprocessing and Analysis}

Two Butterworth $5^{th}$-order low--~and~high--pass filters are subsequently applied with cutoff frequencies at $0.5H$z and $20$Hz respectively to remove low frequency and DC-shift interferences. The low-pass filtering removes possible muscle movement frequency artifacts. The preprocessed EEG signals are next segmented to create event related epochs starting at $-100$ms from the stimulus onset and ending at $800$ms after it.

\subsection{EEG Preprocessing to Clean the Signals from Eye-Movements with Multivariate Empirical Mode Decomposition}

There are many signal processing methods trying to remove eye-- and muscle--movement interferences from EEG signals. The most successful approaches belong the new developments based on empirical mode decomposition~\cite{icassp2009tomek,icassp2012khademul}, since they outperform ICA and blind source separation based approaches~\cite{tomekNEUROCOMPUTING2012}.
There exist already multiple signal processing approaches that extend the classical EMD~\cite{art:emdPAPER} concept which constitutes the fully data adaptive technique to decompose any non--linear or non--stationery signal into a finite set of band-limited basis functions called intrinsic mode functions (IMFs). Each of the decomposed IMFs carry the both amplitude and frequency modulated oscillatory components. The very recent development in the field is the \emph{multivariate EMD} (MEMD)~\cite{MEMDroyalSOC2010}, which is a more generalized extension of the classical univariate EMD technique. The most interesting part of that approach is that it can deal instantaneously with multivariate dataset, like EEG for example. The major novelty proposed in~\cite{MEMDroyalSOC2010} is based on possibility to process the $n$-dimensional signals in the same number of spaces. As a result at each decomposition step the $n$-dimensional envelopes are generated by taking signal projections along different directions in the same dimensional spaces. Next, as in classical univariate EMD, the local mean is calculated as an approximation of the integral of all envelopes along multiple directions in the $n$-dimensional space. The authors in~\cite{MEMDroyalSOC2010} very elegantly resolve a problem related to a lack of formal definition of maxima and minima in multidimensional domains by utilizing a sampling concept based on low discrepancy Hammersley sequence to generate the projections. In a next step, the extrema could be interpolated using cubic splines. The resulting signal envelopes are obtained from the projections along different directions in multidimensional spaces. They are finally averaged to obtain the local mean of the original multivariate EEG signal, as in our case. The whole MEMD decomposition procedure of an input multivariate signal $s(t)$ into a multidimensional set of IMFs, as in~\cite{MEMDieee2011}, could be summarized as follows:
\begin{enumerate}
	\item As the first step, generate the point-set based on the Hammersley sequence for an uniform sampling of the EEG on an $(n-1)-dimensional-sphere$;
	\item In the second step, calculate a projection $\{ p^{\theta_k}(t)\}^T_{t=1}$
	 of the input EEG $\left\{s(t)\right\}^T_{t=1}$ along the direction vector $X^{\theta_k}$, for all $k$ (the whole set of direction vectors) resulting in $\{ p^{\theta_k}(t)\}^K_{k=1}$ set of projections;
	\item In the third step, find the time points $\{ t^{\theta_k}_i\}^K_{k=1}$ representing the maxima in the set of projected signals $\{ p^{\theta_k}(t)\}^K_{k=1}$;
	\item As the fourth step, perform interpolation $[t_i^{\theta_k}, s(t_i^{\theta_k})]$, for all values of $k$, in order to obtain the multivariate envelope curves $\{ v^{\theta_k}(t)\}^K_{k=1}$;
%%%	
%%%	
	\item In the fifth step, for a set of $k$ direction vectors, calculate the mean $m(t)$ of the envelope curves as:
		\begin{equation}
			m(t) = \frac{1}{k}\sum^K_{k=1}v^{\theta_k}(t)\label{eq:MEMDenvelope}
		\end{equation}
	\item Finally, extract/sift the ``detail'' $d(t)$ using $d(t) = X(t) - m(t)$. Similarly as in classical univariate EMD approach, the ``detail'' $d(t)$ shall fulfill the stoppage criterion for a multivariate IMF, so apply the above procedure to $X(t)-d(t)$, otherwise apply it to $d(t)$.
\end{enumerate}
Once the first multivariate IMF is identified, it is subtracted from the input signal and the same process is applied to the resulting signal yielding the second IMF, etc. In the multivariate case, similarly as in univariate one, the residue corresponds to a signal whose projections do not contain enough extrema to form a meaningful multivariate envelope. The stopping criterion in MEMD is also similar to standard EMD~\cite{art:emdPAPER} one, only with a difference that the condition for equality of the number of extrema and zero crossings is not imposed, as extrema could not be properly defined for multivariate signals. The resulting from MEMD decomposition filter banks represent an array of band-pass filters designed to filter, with possible spectral overlap, the input multivariate signal into different frequency bands. The frequency bands are the same for all IMFs which is a very interesting feature of the MEMD method. In case of sequential implementation of classical univariate EMD, each channel would result with different number of IMFs, which would increase a complexity of artifacts identification. The MEMD decomposition allows for simple comparison of the adaptive subband filtering results and rejection only of the contaminated by eye/muscle-movement IMFs.

\begin{figure}[H]
	\begin{center}
	\includegraphics[width=0.7\linewidth]{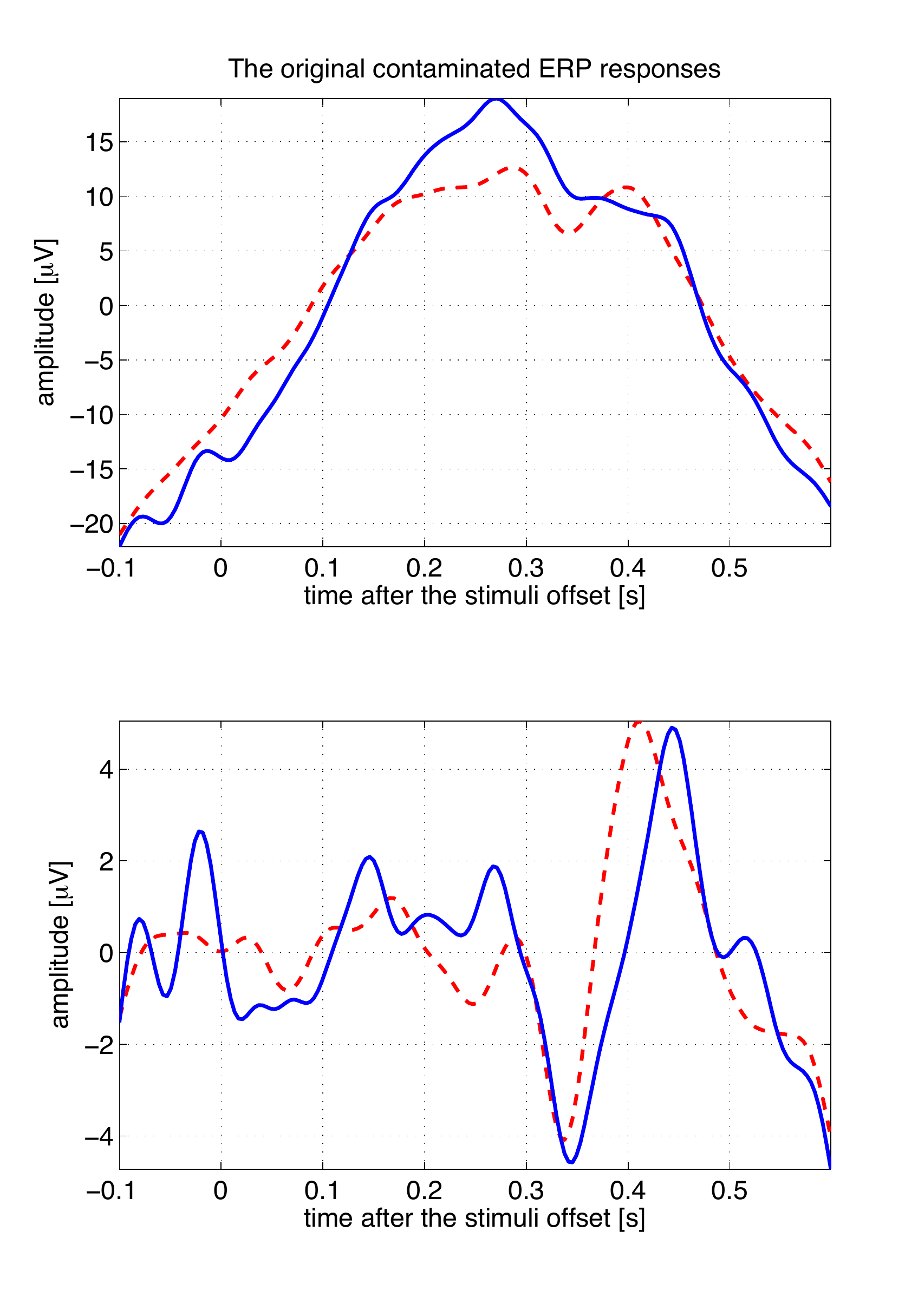}
	\end{center}
	\caption{An example of EMD ERP cleaning. The top panel depicts the original contaminated EEG ERP response with large drifts for both the in \emph{target} (blue, solid line) and \emph{non-target} responses (red, dashed line). The cleaned version is shown in lower panel, where there ERP related peaks are clearly visible.}
	\label{fig:EMD}
\end{figure}

For the eye/muscle-movement identification we apply an $peak-to-peak$ thresholding over the the all IMFs. Those IMFs which exceed $20\mu$V peak--to--peak value are marked as ``contaminated'' and removed from the following reconstruction. The result of the procedure is shown in Figure~\ref{fig:EMD}, where the very strong eye-blinks, completely damaging the averaged ERPs, are successfully removed with the proposed method.

\subsection{The Offline saBCI EEG Analysis Results}

\begin{figure}[H]
	\begin{center}
	\includegraphics[width=0.8\linewidth]{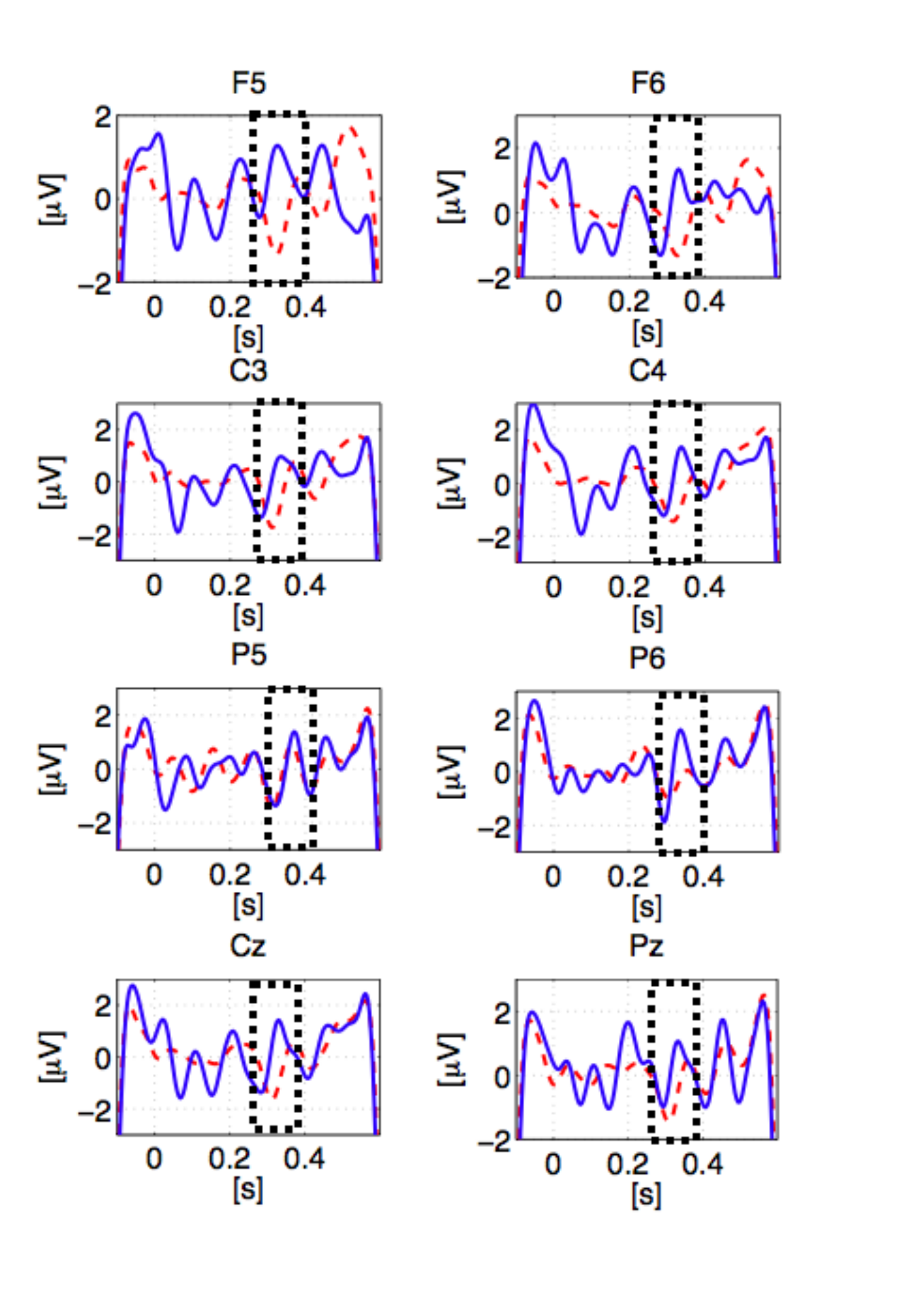}
	\end{center}
	\caption{EEG evoked responses to real sound stimuli. The blue--solid lines represent the averaged responses to the \emph{target} sounds, while the red--dashed lines to the averaged \emph{non-targets} with closer depth setting ($\sim70$dB). The $P300$--response latencies are marked with black--dotted line squares for each electrode.}
	\label{fig:REAL_LOU}
\end{figure}

\begin{figure}[H]
	\begin{center}
	\includegraphics[width=0.8\linewidth]{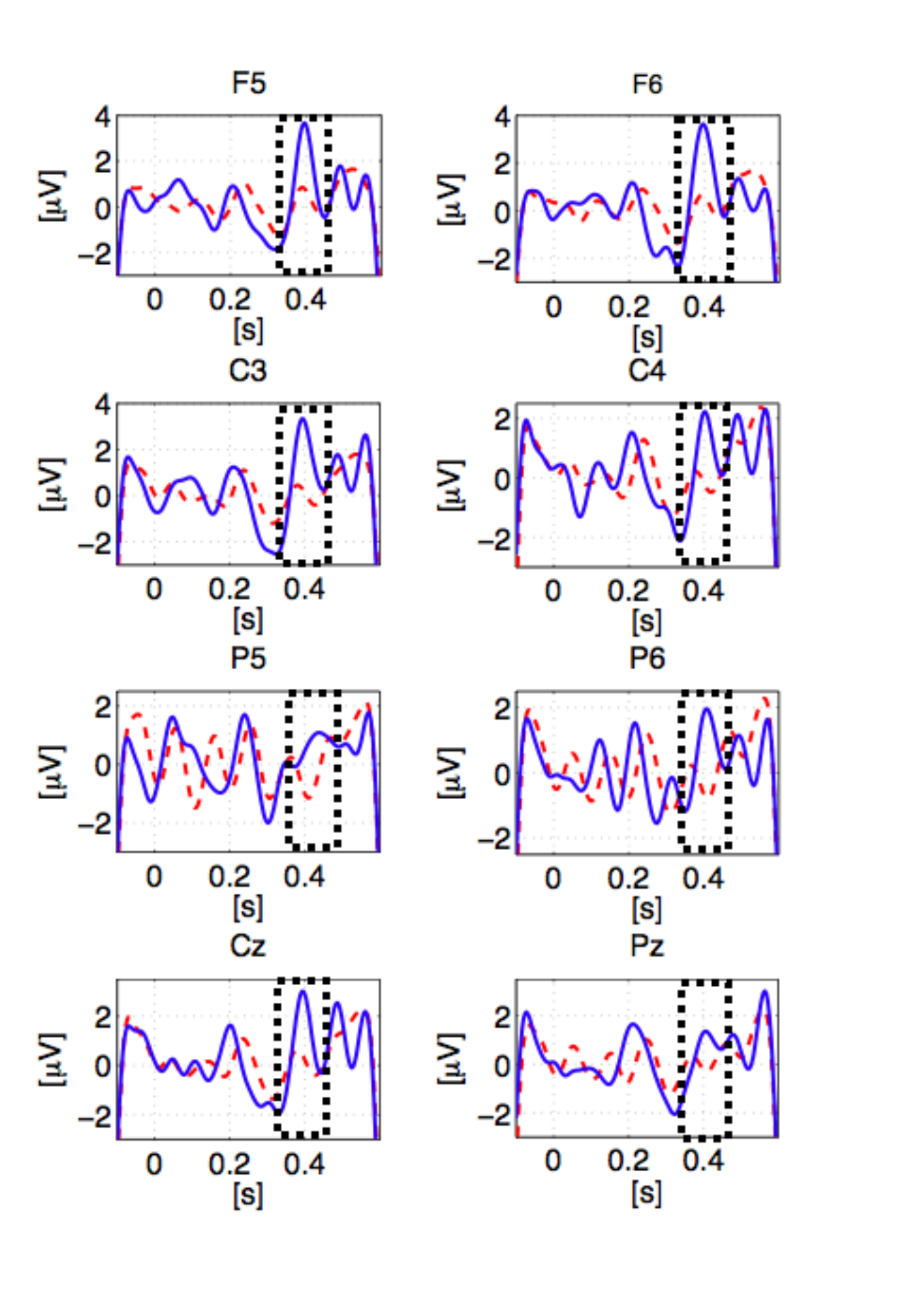}
	\end{center}
	\caption{EEG evoked responses to real sound stimuli. The blue--solid lines represent the averaged responses to the \emph{target} sounds, while the red--dashed lines to the averaged \emph{non-targets} with farther depth setting ($\sim30$dB). The $P300$--response latencies are marked with black--dotted line squares for each electrode.}
	\label{fig:REAL_SML}
\end{figure}

\begin{figure}[H]
	\begin{center}
	\includegraphics[width=0.8\linewidth]{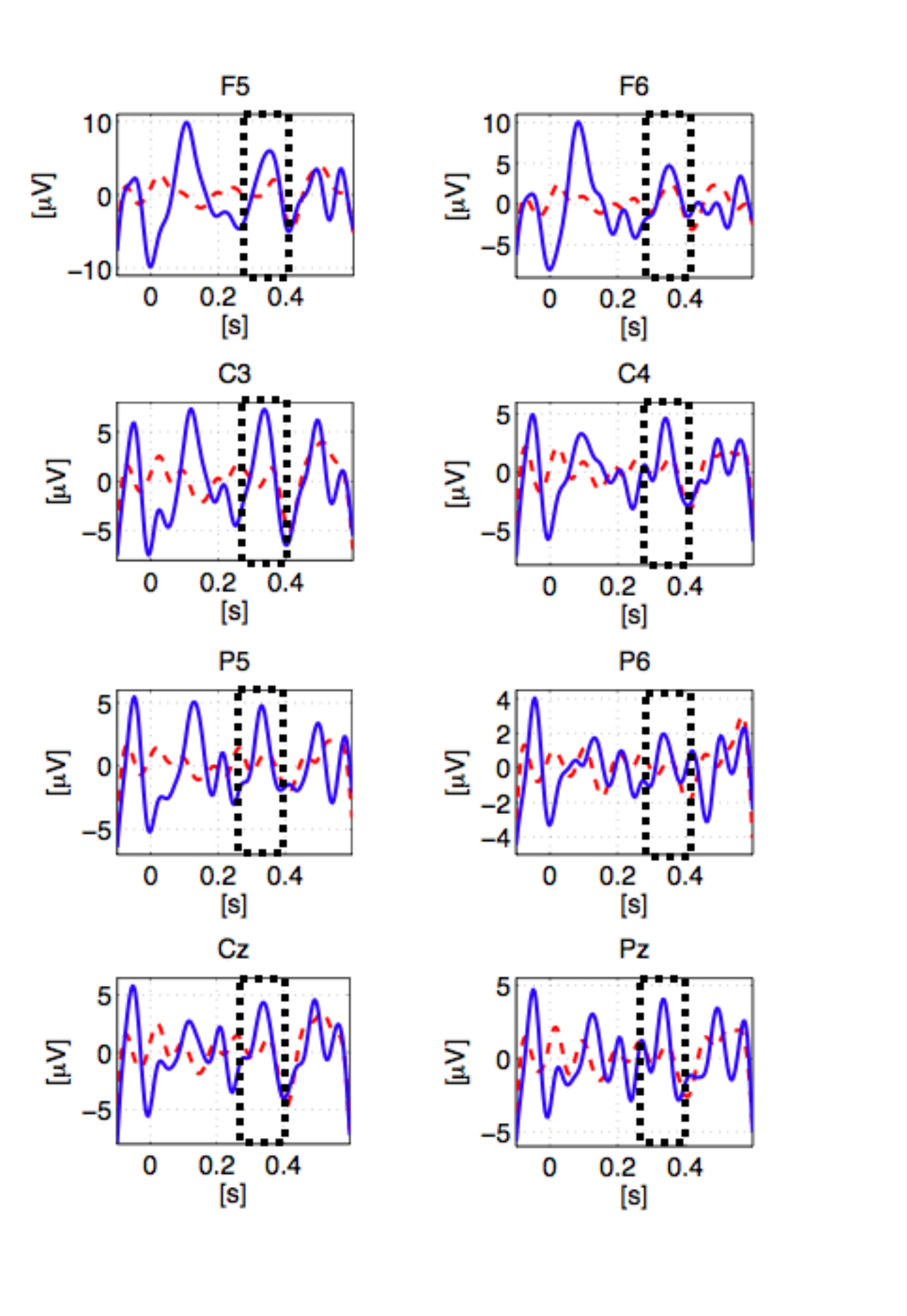}
	\end{center}
	\caption{EEG evoked responses to virtual sound stimuli using VBAP. The blue--solid lines represent the averaged responses to the \emph{target} sounds, while the red--dashed lines to the averaged \emph{non-targets} with farther depth setting ($\sim70$dB). The $P300$--response latencies are marked with black--dotted line squares for each electrode.}
	\label{fig:VIR_LOU}
\end{figure}

\begin{figure}[H]
	\begin{center}
	\includegraphics[width=0.8\linewidth]{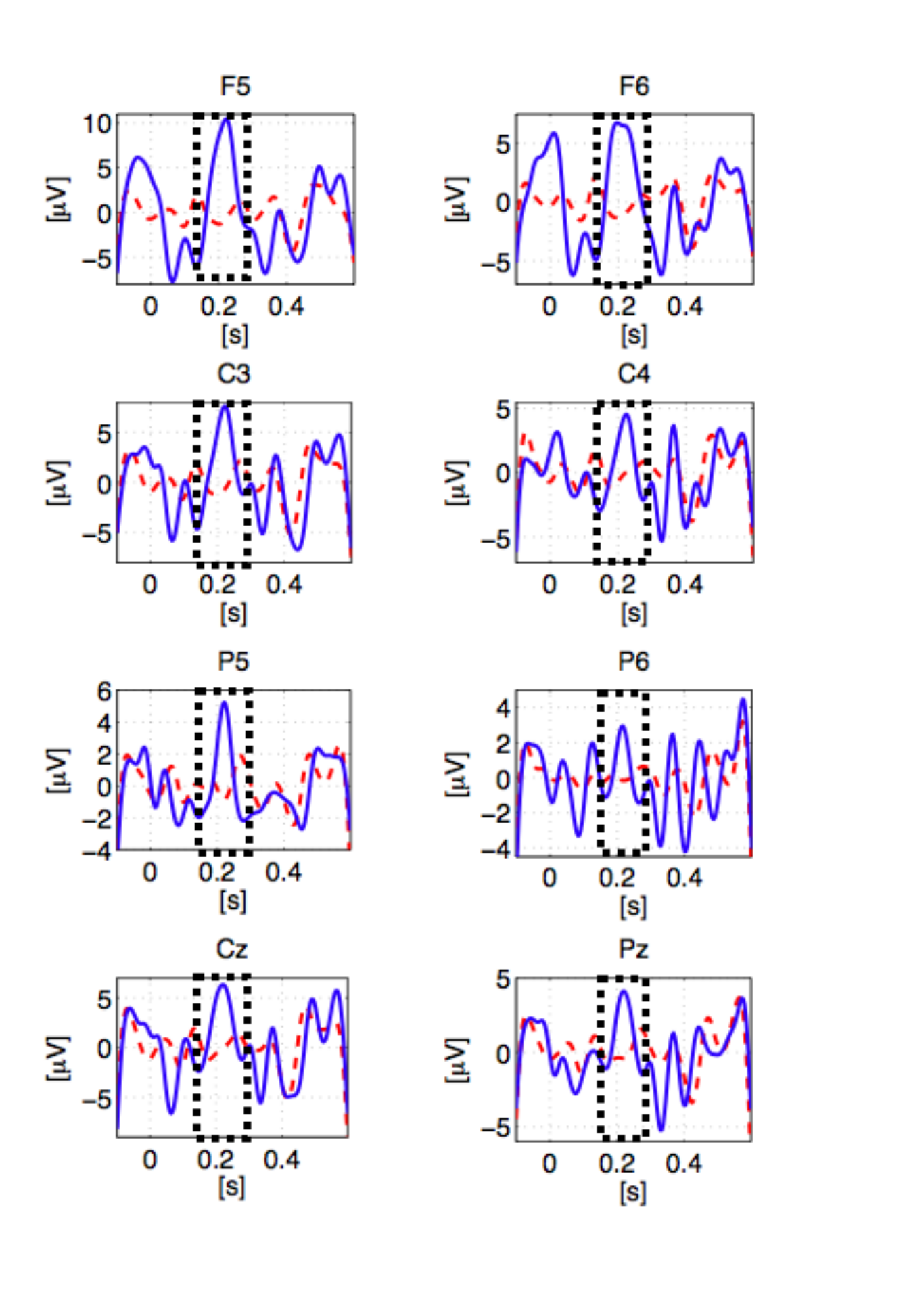}
	\end{center}
	\caption{EEG evoked responses to real sound stimuli. The blue--solid lines represent the averaged responses to the \emph{target} sounds, while the red--dashed lines to the averaged \emph{non-targets} with farther depth setting ($\sim30$dB). The $P300$--response latencies are marked with black--dotted line squares for each electrode.}
	\label{fig:VIR_SML}
\end{figure}

The averaged results of the conducted EEG experiments with the six subjects in the offline saBCI mode are presented in Figures~\ref{fig:REAL_LOU}--\ref{fig:VIR_SML}. Each of those figures has eight panels comparing the averaged responses to \emph{targets} (solid--blue lines) and to \emph{non--targets} (dashed--red lines) for each EEG electrode used in our experiments. 
The presented ERPs are the grand averages for all subjects and all sound timbres used in our experiments. The comparison between the real (see Figures~\ref{fig:REAL_LOU}~and~\ref{fig:REAL_SML}) and virtual (see Figures~\ref{fig:VIR_LOU}~and~\ref{fig:VIR_SML}) sound images show that strong $P300$ responses (marked by black dotted--line squares) could be obtained in both cases. This results show that the both real and virtual sound reproduction systems yield similar results of saBCI stimuli generation. Additionally we can see from the results with farther sound (smaller loudness) in Figures~\ref{fig:REAL_SML}~and~\ref{fig:VIR_SML} that the $P300$ response seems to be stronger in amplitude comparing to closer distance (larger loudness) results in Figures~\ref{fig:REAL_LOU}~and~\ref{fig:VIR_LOU}. This very interesting phenomenon will be a target of our future research. The most important message from our current experimental session is that the proposed simplified VBAP application is a valid method to create the good saBCI stimuli. We could not observe any dependence of the spatial sound locations on $P300$ responses which is also a good result for future saBCI applications.

\section{Conclusions}

The purpose of the presented research was to check whether virtual sound images creation procedures were useful for the saBCI paradigm development. We have been researching a possibility to increase the possible number of BCI commands for future applications by keeping a limited number of the loudspeaker units. We conducted two experimental series to test the hypotheses of real and virtual images equality for saBCI paradigm design, sound depth manipulation, and whether the various timbres modulate the psychophysical and brain responses.

The very small, yet not significant, differences in psychophysical experiments were later not reproduced in EEG experiments, in offline saBCI setting, which is a very good sign for the future developments in saBCI research field. The psychophysical experiments confirmed, based on the lack of significant differences of pairwise statistical tests, that the directions and sound reproduction systems (real or virtual) result with the same behavioral responses.

The brain responses in offline saBCI setting further confirmed the original hypothesis of the similarity of real and virtual sound settings. We reported also on the very interesting ERP modulation depending on sound depth (loudness) which resulted with slightly different, larger for the farther sound, $P300$ responses. This topic will be further investigated in our future research.

In the future, we will develop and online saBCI application based on the proposed simplified VBAP approach in order to create a more easy to use interfacing paradigm with possible larger and independent from the loudspeakers quantity the multiple commands.  

Additionally we plan to further expend the system to use fully three--dimensional sound field by adding elevation variability, which shall further improve and enhance the system capability. Vertical sound localization is a bit more difficult to reproduce, yet it shall improve the interfacing comfort.

The presented results are a step forward in development of vision-free and auditory spatial brain computer interfacing paradigm which shall in the future support and improve comfort of life of locked--in patients.

\section*{Acknowledgments}

%First and foremost, we would like to thank God for being my strength and guide in the writing of this thesis.

This research was supported in part by the Strategic Information and Communications R\&D Promotion Programme no. 121803027 of The Ministry of Internal Affairs and Communication in Japan, and by KAKENHI, the Japan Society for the Promotion of Science grant no. 12010738.

\bibliographystyle{ieeetr}
\bibliography{nozomuAPSIPA2012}

\newpage
\section*{Figure Legends}

\begin{description}
	\item[Figure~\ref{fig:AuditoryBCI}] Schematic drawing of the proposed auditory BCI system. The subject in the center is surrounded by horizontally distributed loudspeakers. EEG signals are processed by the BCI processing stages in synchrony with presented spatial auditory stimuli.
	\item[Figure~\ref{fig:P300}] An example of averaged brain evoked responses to \emph{target} (solid blue line) and \emph{non-target} (dotted red line). The difference in ERP waves is mostly pronounced around $300$ms where usually $P300$\emph{/aha-response} occurs. The zero second time point indicates stimulus onset.
	\item[Figure~\ref{fig:vbap}] A schematic figure of the VBAP method implementation.
	\item[Figure~\ref{fig:2ch_vbap}] A concept of the simplified horizontal panning of sound source with two loudspeakers based on the original VBAP~\cite{pulkkiVBAP1997} implementation.
	\item[Figure~\ref{fig:SpeakerLocation}] The locations of the eight loudspeakers in our experiments. The frontal to the subject loudspeaker at a position of $0^\circ$.
	\item[Figure~\ref{fig:EMD}] An example of EMD ERP cleaning. The top panel depicts the original contaminated EEG ERP response with large drifts for both the in \emph{target} (blue, solid line) and \emph{non-target} responses (red, dashed line). The cleaned version is shown in lower panel, where there ERP related peaks are clearly visible.
	\item[Figure~\ref{fig:REAL_LOU}] EEG evoked responses to real sound stimuli. The blue--solid lines represent the averaged responses to the \emph{target} sounds, while the red--dashed lines to the averaged \emph{non-targets} with closer depth setting ($\sim70$dB). The $P300$--response latencies are marked with black--dotted line squares for each electrode.
	\item[Figure~\ref{fig:REAL_SML}] EEG evoked responses to real sound stimuli. The blue--solid lines represent the averaged responses to the \emph{target} sounds, while the red--dashed lines to the averaged \emph{non-targets} with farther depth setting ($\sim30$dB). The $P300$--response latencies are marked with black--dotted line squares for each electrode.
	\item[Figure~\ref{fig:VIR_LOU}] EEG evoked responses to virtual sound stimuli using VBAP. The blue--solid lines represent the averaged responses to the \emph{target} sounds, while the red--dashed lines to the averaged \emph{non-targets} with farther depth setting ($\sim70$dB). The $P300$--response latencies are marked with black--dotted line squares for each electrode.
	\item[Figure~\ref{fig:VIR_SML}] EEG evoked responses to real sound stimuli. The blue--solid lines represent the averaged responses to the \emph{target} sounds, while the red--dashed lines to the averaged \emph{non-targets} with farther depth setting ($\sim30$dB). The $P300$--response latencies are marked with black--dotted line squares for each electrode.
\end{description}

\newpage
\section*{Tables}

\begin{table}[H]
\caption{The psychophysical experiment conditions.}
\label{tb:BPCondition}
\begin{center}
\begin{tabular}
{|l|l|}
\hline
Condition & Parameters \\ \hline\hline
Sound stimuli & white noise, MIDI sound, effect sound\\ \hline
Sound source & eight channel loudspeaker system \\ \hline
Response input &  computer keyboard \\ \hline
Sound directions & 8 \\
& ($-135^\circ$, $-90^\circ$, $-45^\circ$, $0^\circ$, $45^\circ$, $90^\circ$, $135^\circ$, $180^\circ$) \\ \hline
Volume levels & $\sim30$dB and $\sim70$dB \\ 
\hline
Subjects & $6$ \\ \hline
Ttrials & $12~(2~\mbox{trials} \times 3~\mbox{stimuli} \times 2~\mbox{real/virtual-sounds})$ \\ \hline
Stimulus length & $500$ms \\ \hline
SOA & $1000$ms \\ \hline
\end{tabular}
\end{center}
\end{table}

\begin{table}[H]
\caption{Accuracy rate of psychophysical response to real sound stimuli with the nearer sound source setting ($\sim70$dB) as a function of various stimuli (AR = accuracy rate, ART = average response time, $\sigma$ = standard deviation).}
\label{tb:REAL_LOU_sti}
\begin{center}
\begin{tabular}{|l|c|c|c|}
\hline
Stimulus type & AR~[\%] & ART~[ms] & $\sigma$~[ms]\\ 
\hline\hline
White noise & $91.0$ & $574$ & $103$\\ \hline
MIDsound & $93.2$ & $571$ & $102$\\ \hline
Effect sound & $82.0$ & $570$ & $104$\\
\hline
\end{tabular}
\end{center}
\end{table}

\begin{table}[H]
\caption{Accuracy rate of psychophysical response to real sound stimuli with the nearer sound source setting ($\sim70$dB) as a function of the spatial direction (AR = accuracy rate, ART = average response time, $\sigma$ = standard deviation).}
\label{tb:REAL_LOU_dir}
\begin{center}
\begin{tabular}{|c|c|c|c|}
\hline
Sound directions & AR~[\%] & ART~[ms] & $\sigma$~[ms]\\ \hline\hline
$-135^\circ$ & $94.4$ & $603$ & $91.6$\\ \hline
$-90^\circ$ & $91.0$ & $573$ & $98.4$\\ \hline
$-45^\circ$ & $94.4$ & $554$ & $79.5$\\ \hline
$0^\circ$ & $91.6$ & $535$ & $94.2$\\ \hline
$45^\circ$ & $100.0$ & $547$ & $88.9$\\ \hline
$90^\circ$ & $88.8$ & $582$ & $95.7$\\ \hline
$135^\circ$ & $88.8$ & $590$ & $102.0$\\ \hline
$180^\circ$ & $86.1$ & $537$ & $128.0$\\
\hline
\end{tabular}
\end{center}
\end{table}

\begin{table}[H]
\caption{Accuracy rate of psychophysical response to real sound stimuli with the nearer sound source setting ($\sim30$dB) as a function of various stimuli (AR = accuracy rate, ART = average response time, $\sigma$ = standard deviation).}
\label{tb:REAL_SML_sti}
\begin{center}
\begin{tabular}{|l|c|c|c|}
\hline
Stimulus type & AR~[\%] & ART~[ms] & $\sigma$~[ms]\\ 
\hline\hline
White noise & $90.6$ & $582$ & $117$\\ \hline
MIDsound & $91.7$ & $539$ & $105$\\ \hline
Effect sound & $75.0$ & $562$ & $99.4$\\
\hline
\end{tabular}
\end{center}
\end{table}

\begin{table}[H]
\caption{Accuracy rate of psychophysical response to real sound stimuli with the nearer sound source setting ($\sim30$dB) as a function of the spatial direction (AR = accuracy rate, ART = average response time, $\sigma$ = standard deviation).}
\label{tb:REAL_SML_dir}
\begin{center}
\begin{tabular}{|c|c|c|c|}
\hline
Sound directions & AR~[\%] & ART~[ms] & $\sigma$~[ms]\\ \hline\hline
$-135^\circ$ & $91.6$ & $562$ & $113$\\ \hline
$-90^\circ$ & $91.6$ & $531$ & $90.2$\\ \hline
$-45^\circ$ & $83.3$ & $553$ & $84.7$\\ \hline
$0^\circ$ & $80.5$ & $541$ & $115$\\ \hline
$45^\circ$ & $88.8$ & $531$ & $92.9$\\ \hline
$90^\circ$ & $77.7$ & $589$ & $98.0$\\ \hline
$135^\circ$ & $88.8$ & $617$ & $148$\\ \hline
$180^\circ$ & $83.3$ & $562$ & $98.9$\\
\hline
\end{tabular}
\end{center}
\end{table}

\begin{table}[H]
\caption{Accuracy rate of psychophysical response to virtual sound stimuli with the nearer sound source setting ($\sim70$dB) as a function of various stimuli (AR = accuracy rate, ART = average response time, $\sigma$ = standard deviation).}
\label{tb:VIR_LOU_sti}
\begin{center}
\begin{tabular}{|l|c|c|c|}
\hline
Stimulus type & AR~[\%] & ART~[ms] & $\sigma$~[ms]\\ 
\hline\hline
White noise & $54.6$ & $612$ & $115$\\ \hline
MIDsound & $44.7$ & $571$ & $115$\\ \hline
Effect sound & $47.3$ & $623$ & $113$\\
\hline
\end{tabular}
\end{center}
\end{table}

\begin{table}[H]
\caption{Accuracy rate of psychophysical response to virtual sound stimuli with the nearer sound source setting ($\sim70$dB) as a function of the spatial direction (AR = accuracy rate, ART = average response time, $\sigma$ = standard deviation).}
\label{tb:VIR_LOU_dir}
\begin{center}
\begin{tabular}{|c|c|c|c|}
\hline
Sound directions & AR~[\%] & ART~[ms] & $\sigma$~[ms]\\ \hline\hline
$-135^\circ$ & $67.2$ & $630$ & $135$\\ \hline
$-90^\circ$ & $47.7$ & $641$ & $92$\\ \hline
$-45^\circ$ & $49.4$ & $595$ & $149$\\ \hline
$0^\circ$ & $56.6$ & $583$ & $103$\\ \hline
$45^\circ$ & $43.8$ & $634$ & $125$\\ \hline
$90^\circ$ & $44.4$ & $619$ & $106$\\ \hline
$135^\circ$ & $45.0$ & $605$ & $104$\\ \hline
$180^\circ$ & $71.6$ & $611$ & $142$\\
\hline
\end{tabular}
\end{center}
\end{table}

\begin{table}[H]
\caption{Accuracy rate of psychophysical response to virtual sound stimuli with the nearer sound source setting ($\sim30$dB) as a function of various stimuli (AR = accuracy rate, ART = average response time, $\sigma$ = standard deviation).}
\label{tb:VIR_SML_sti}
\begin{center}
\begin{tabular}{|l|c|c|c|}
\hline
Stimulus type & AR~[\%] & ART~[ms] & $\sigma$~[ms]\\ 
\hline\hline
White noise & $46.9$ & $603$ & $109$\\ \hline
MIDsound & $44.3$ & $607$ & $132$\\ \hline
Effect sound & $45.6$ & $625$ & $90.9$\\
\hline
\end{tabular}
\end{center}
\end{table}

\begin{table}[H]
\caption{Accuracy rate of psychophysical response to virtual sound stimuli with the nearer sound source setting ($\sim30$dB) as a function of the spatial direction (AR = accuracy rate, ART = average response time, $\sigma$ = standard deviation).}
\label{tb:VIR_SML_dir}
\begin{center}
\begin{tabular}{|c|c|c|c|}
\hline
Sound directions & AR~[\%] & ART~[ms] & $\sigma$~[ms]\\ \hline\hline
$-135^\circ$ & $40.0$ & $603$ & $78.1$\\ \hline
$-90^\circ$ & $49.4$ & $605$ & $108$\\ \hline
$-45^\circ$ & $51.6$ & $597$ & $127$\\ \hline
$0^\circ$ & $37.7$ & $586$ & $121$\\ \hline
$45^\circ$ & $37.2$ & $668$ & $98.9$\\ \hline
$90^\circ$ & $40.5$ & $607$ & $111$\\ \hline
$135^\circ$ & $42.2$ & $648$ & $126$\\ \hline
$180^\circ$ & $65.5$ & $620$ & $82.1$\\
\hline
\end{tabular}
\end{center}
\end{table}

\begin{table}[H]
\caption{The offline saBCI EEG experimental conditions.}
\label{tb:Final_Condition}
\begin{center}
\begin{tabular}
{|l|l|}
\hline
Condition & Parameters \\ \hline\hline
Sound stimuli & white noise, MIDI sound, effect sound\\ \hline
Sound source &  eight channel loudspeaker system \\ \hline
Response input &  8-electrodes EEG system \\ \hline
Sound directions & 8 \\
& ($-135^\circ$, $-90^\circ$, $-45^\circ$, $0^\circ$, $45^\circ$, $90^\circ$, $135^\circ$, $180^\circ$) \\ \hline
Volume levels & $\sim30$dB and $\sim70$dB \\ 
\hline
Subjects & $6$ \\ \hline
Ttrials & $12~(2~\mbox{trials} \times 3~\mbox{stimuli} \times 2~\mbox{real/virtual-sounds})$ \\ \hline
Stimulus length & $500$ms \\ \hline
SOA & $1000$ms \\ \hline
\end{tabular}
\end{center}
\end{table}

\newpage
\section*{Figures}

\end{document}